\documentstyle[aps,prl,preprint,floats,epsf]{revtex}

\begin{document}

\preprint{\tighten\vbox{\hbox{\hfil CLNS 97/1470}
                        \hbox{\hfil CLEO 97-5}
}}

\title{First Observation of Inclusive $B$ Decays to the Charmed Strange Baryons $\Xi_c^0$ and $\Xi_c^+$}

\author{CLEO Collaboration}
\date{\today}

\maketitle
\tighten

\input{psfig}

\newcommand{\cesr}{CESR}
\newcommand{\cleo}{CLEO}
\newcommand{\cleoii}{CLEO II}
\newcommand{\epem}{\mbox{$e^+e^-$}}
\newcommand{\ufours}{\mbox{$\Upsilon(4S)$}}
\newcommand{\B}{\mbox{$B$}}
\newcommand{\Bbar}{\mbox{$\overline{B}$}}
\newcommand{\BBbar}{\mbox{$B\overline{B}$}}
\newcommand{\ccbar}{\mbox{$c\overline{c}$}}
\newcommand{\dedx}{\mbox{$dE/dx$}}
\newcommand{\tof}{\mbox{\rm TOF\/}}
\newcommand{\mev   }{\mbox{\rm MeV}} 
\newcommand{\mevc  }{\mbox{\rm MeV/$c$}}
\newcommand{\mevcsq}{\mbox{\rm MeV/$c^2$}}
\newcommand{\gev   }{\mbox{\rm GeV}} 
\newcommand{\gevc  }{\mbox{\rm GeV/$c$}}
\newcommand{\gevcsq}{\mbox{\rm GeV/$c^2$}}
\newcommand{\ubar}{\mbox{$\overline{u}$}}
\newcommand{\ubard}{\mbox{$\overline{u}d$}}
\newcommand{\cbar}{\mbox{$\overline{c}$}}
\newcommand{\cbars}{\mbox{$\overline{c}s$}}
\newcommand{\p}{\mbox{$p$}}
\newcommand{\pimi}{\mbox{$\pi^-$}}
\newcommand{\pipl}{\mbox{$\pi^+$}}
\newcommand{\Y}{\mbox{$Y$}}
\newcommand{\Ybar}{\mbox{$\overline{Y}$}}
\newcommand{\N}{\mbox{$N$}}
\newcommand{\Nbar}{\mbox{$\overline{N}$}}
\newcommand{\lz}{\mbox{$\Lambda$}}
\newcommand{\cas}{\mbox{$\Xi$}}
\newcommand{\casm}{\mbox{$\Xi^-$}}
\newcommand{\casc}{\mbox{$\Xi_c$}}
\newcommand{\cascz}{\mbox{$\Xi^0_c$}}
\newcommand{\cascp}{\mbox{$\Xi^+_c$}}
\newcommand{\nul}{\mbox{$\nu_l$}}
\newcommand{\lc}{\mbox{$\Lambda_c$}}
\newcommand{\x}{\mbox{$X$}}
\newcommand{\X}{\mbox{$X$}}
\newcommand{\xp}{\mbox{$x_p$}} 
\newcommand{\mcc}[2]{\multicolumn{#1}{c}{#2}}
\def\plb#1#2#3#4{#1, Phys.\ Lett.\ B {\bf#2}, #3 (#4)}
\def\prd#1#2#3#4{#1, Phys.\ Rev.\ D {\bf#2}, #3 (#4)}
\def\prl#1#2#3#4{#1, Phys.\ Rev.\ Lett.\ {\bf#2}, #3 (#4)}
\def\etal{{\em et al.}}

\begin{abstract} 
% Insert abstract here.

Using data collected in the region of the \ufours\
resonance with the \cleoii\ detector operating at the 
Cornell Electron Storage Ring (\cesr ), 
we present the first observation of \B\ mesons decaying into 
the charmed strange baryons \cascz\ and \cascp.
We find $79\pm 27$ \cascz\ and $125\pm 28$ \cascp\ candidates 
from \B\ decays, leading to product branching fractions of
${\cal B}(\overline{B}\to\cascz\X){\cal B}(\cascz\to\casm\pipl) = 
(0.144\pm 0.048\pm 0.021)\times 10^{-3}$ and 
${\cal B}(\overline{B}\to\cascp\X){\cal B}(\cascp\to\casm\pipl\pipl) = 
(0.453\pm 0.096\:^{+\:0.085}_{-\:0.065})\times 10^{-3}$.

\end{abstract}
\newpage

{
\renewcommand{\thefootnote}{\fnsymbol{footnote}}

% Insert author and address list here

\begin{center}
B.~Barish,$^{1}$ M.~Chadha,$^{1}$ S.~Chan,$^{1}$ G.~Eigen,$^{1}$
J.~S.~Miller,$^{1}$ C.~O'Grady,$^{1}$ M.~Schmidtler,$^{1}$
J.~Urheim,$^{1}$ A.~J.~Weinstein,$^{1}$ F.~W\"{u}rthwein,$^{1}$
D.~M.~Asner,$^{2}$
% ...
\end{center}
 
\small
\begin{center}
$^{1}${California Institute of Technology, Pasadena, California 91125}\\
$^{2}${University of California, San Diego, La Jolla, California 92093}\\
% ...
\end{center}
 
\setcounter{footnote}{0}
}
\newpage

% Insert body of the text here.

Charmed baryon production from the decays of \B\ mesons has been 
previously reported by ARGUS\cite{btobaryonargus} and 
\cleo\cite{btobaryoncleo,btosigmac}. Here, we report the first observation of 
the charmed-strange baryons \cascz\ and \cascp\ from \B\ decays\cite{conj},
which have previously been observed only in direct charm 
production\cite{biagi,coteus,alam,avery,barlag,albrecht}. 

\begin{figure}
\vspace*{0.5in}
\mbox{\psfig{figure=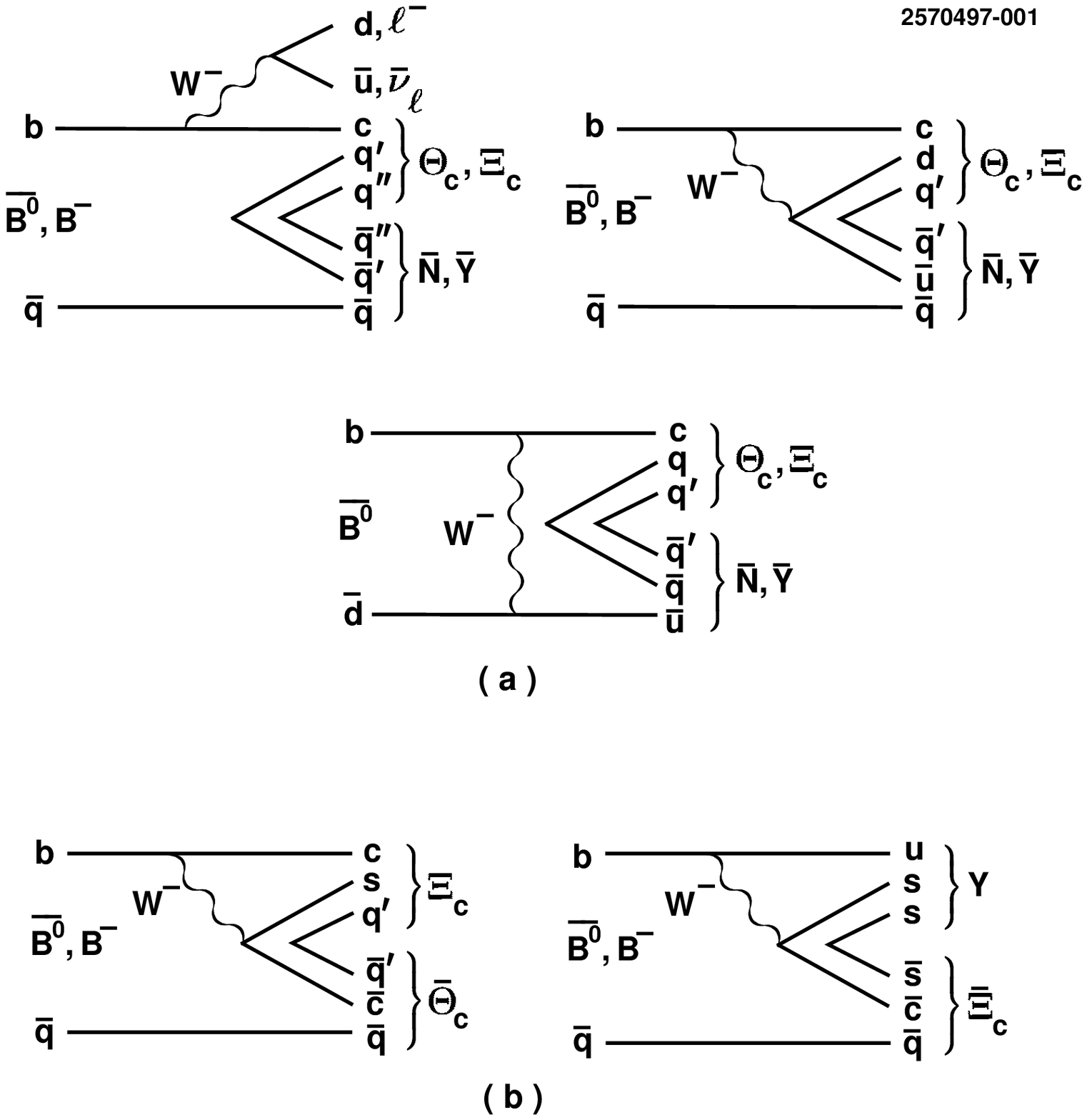}}
\vspace*{0.25in}
\caption
{
Possible $\B \to$ baryon decay mechanisms:
(a) $\Bbar \to \Theta_c\Nbar\x$ and $\casc\Ybar\x$ ,
(b) $\Bbar \to \casc\overline{\Theta}_c\x$ and
    $\Bbar \to \Y\overline{\Xi}_c\x$;
$N$ stands for any non-strange non-charmed baryon,
$Y$ for any strange and non-charmed baryon, and
$\Theta_c$ for any charmed and non-strange baryon.
}
\label{fig:bbaryon_feyn}
\end{figure}

In \epem\ annihilations at the \ufours\ resonance 
(10.58 GeV), charmed baryons can be produced either from 
$B$ meson decay or from hadronization of \ccbar\ quarks 
produced in the continuum.
Since the $b$ quark couples predominantly to the $c$ quark, $B$ meson decays 
to the charmed strange baryons \cascz\ $(csd)$ and \cascp\ $(csu)$ will 
proceed through either spectator or exchange diagrams.
Decays mediated by the coupling $b \to c W^-$ with $W^- \to \ubar d$ produce 
final states of the form $\casc \Ybar X_h$ and $\casc \Nbar X_s$, where 
$Y$ is a hyperon ($\Lambda$, $\Sigma$, $\Xi$, etc.), $N$ is a 
nucleon, and $X_h(X_s)$ denotes non-strange (strange) multi-body 
mesonic states (see Figure~\ref{fig:bbaryon_feyn}(a)).
As shown in Figure~\ref{fig:bbaryon_feyn}(b), decays mediated by 
$b \to c W^-$ with $W^- \to \cbar s$ can lead to states of the 
form $\casc \overline{\Theta}_c$\cite{dunietz,ammar}, where $\Theta_c$ 
denotes any charmed non-strange baryon. 
The authors of Refs.\cite{ball} and\cite{chernyak} predict branching ratios of 
$(1.0-1.8)\times 10^{-3}$ for those decays. 
The process $b \to u W^-$ with 
$W^- \to \cbar s$ leads to final states of the form $\overline{\Xi}_c Y$, 
but should be highly suppressed by the small $b \to u$ coupling. 

There are several theoretical calculations that attempt to derive the 
two-body contribution to charmed baryon production in $B$ decays.
In the diquark model\cite{ball} baryons of spin 
$\frac{1}{2}(\frac{3}{2})$ are modeled 
as bound states of quarks and scalar (vector) diquarks.
The $b$ quark decays to a scalar diquark and an antiquark;
the latter combines with the light antiquark accompanying the $b$ quark 
to form an antidiquark. The creation of a $q\bar q$ pair then leads to a 
baryon and antibaryon in the final state.
The authors of Ref.\cite{chernyak} calculate decay amplitudes based on QCD sum 
rules, replacing both the $B$ meson and the charmed baryon in the final state
by suitable interpolating currents.
There are also treatments that determine the rates for exclusive baryonic
$B$ decays in terms of three reduced matrix elements\cite{savage}, 
on the basis of the quark diagram scheme\cite{kohara}, using the 
constituent quark model\cite{korner}, and using the pole model\cite{jarfi}. 
The latter four calculations do not quote explicit predictions for 
branching fractions of $B$ decay modes which yield \casc\ baryons. 

For this analysis we used 3.1~{\rm fb$^{-1}$} of data taken on the
\ufours\ resonance, corresponding to 3.3 million \BBbar\ events.
To estimate and subtract continuum background, 1.6~{\rm fb$^{-1}$} 
of data were collected 60 \mev\ below the resonance.
The data were collected with the \cleoii\ detector operating at 
the Cornell Electron Storage Ring, \cesr. 
The CLEO II detector\cite{kubota} is a general purpose solenoidal-magnet 
detector with excellent charged particle and 
shower energy detection capabilities. The detector consists of a charged
particle tracking system surrounded by a scintillation counter time-of-flight
 system and an electromagnetic shower detector consisting of 7800 
thallium-doped cesium iodide crystals.  These detectors are installed within
a 1.5~T superconducting solenoidal magnet.  Incorporated in the return yoke of
the magnet are chambers for muon detection.

Charge measurements from the drift chamber wires provide specific ionization
loss $(dE/dx)$ information.  
To obtain hadron identification, $dE/dx$ and available time-of-flight (TOF) 
measurements are combined to define a joint 
$ \chi_i^2 = [ \{ (dE/dx)_{\rm meas} - (dE/dx)_{\rm exp} \} / 
\sigma_{dE/dx} ]_i^2 + 
[ \{ (T)_{\rm meas} - (T)_{\rm exp} \} / \sigma_{\rm TOF} ]_i^2 $ , where $i$ 
corresponds to the pion, kaon, and proton hypotheses.
A $\chi^2$-probability is then calculated for each hypothesis, and 
particle identification levels for each of the hypotheses are derived 
by normalizing to the sum of the three probabilities.
A particle is identified with a specific hypothesis if its particle  
identification level for it is greater than 0.05.

We reconstruct \cascz\ (\cascp) candidates through the decay chain 
$\cascz \to \casm\pipl$ ($\cascp \to \casm\pipl\pipl$), 
$\casm \to \lz\pimi$, and $\lz \to \p\pimi$. 
We study the \casc\ momentum spectra using the scaled momentum 
$\xp \equiv p/(E^2_{\rm beam} - m^2_{\Xi_c})^{1/2}$, 
where $p$ and $m_{\Xi_c}$ are the \casc\ momentum and mass, respectively, 
and $E_{\rm beam}$ is the beam energy. 
We require $\xp < 0.5$, the kinematic limit for \casc\ baryons produced
from $B$ decays. This requirement reduces the background from continuum \ccbar. 

The \lz\ candidates are formed from pairs of oppositely charged tracks,
assuming the higher momentum track to be a proton and
the lower momentum track to be a pion.  
We also require the higher momentum track to be 
consistent with the proton hypothesis.
The invariant mass of \lz\ candidates has to be within
5.0 \mevcsq\ (corresponding to 2.5 standard deviations) of the known \lz\ mass.
We have not required \lz\ candidates to point towards the primary vertex, 
since \lz's decaying from \casm's can travel as 
much as a few centimeters before decaying and can have appreciable 
impact parameters. 
To reduce the background from tracks coming 
from the interaction point,
we require the radial distance %in the transverse plane 
of the \lz\ decay vertex from the beam line to be greater than 2 mm.
 
The \casm\ candidates are formed by combining each \lz\ candidate with the 
remaining negatively charged tracks in the event, assuming the additional 
track to be a pion. 
The decay vertex of the \casm\ candidate is reconstructed by intersecting 
the extrapolated \lz\ path with the negatively charged track. 
We require the radial distance %in the transverse plane 
of the \casm\ decay vertex from the beam line to be greater than 2 mm and 
less than the radial distance of the \lz\ decay vertex. 
In addition, the reconstructed \casm\ momentum vector has to point back 
to the interaction point.
The invariant mass of the \casm\ candidates has to be within 6.5 \mevcsq\
(corresponding to 3 standard deviations) of the known \casm\ mass.

To reconstruct \cascz\ candidates, 
we form combinations of \casm\ with one positively charged track, and
to reconstruct \cascp\ candidates, 
we combine each \casm\ with two positively charged tracks.
These additional charged tracks are required to originate from the 
interaction point and to be consistent with the pion hypothesis. 

\begin{figure}
\mbox{\psfig{figure=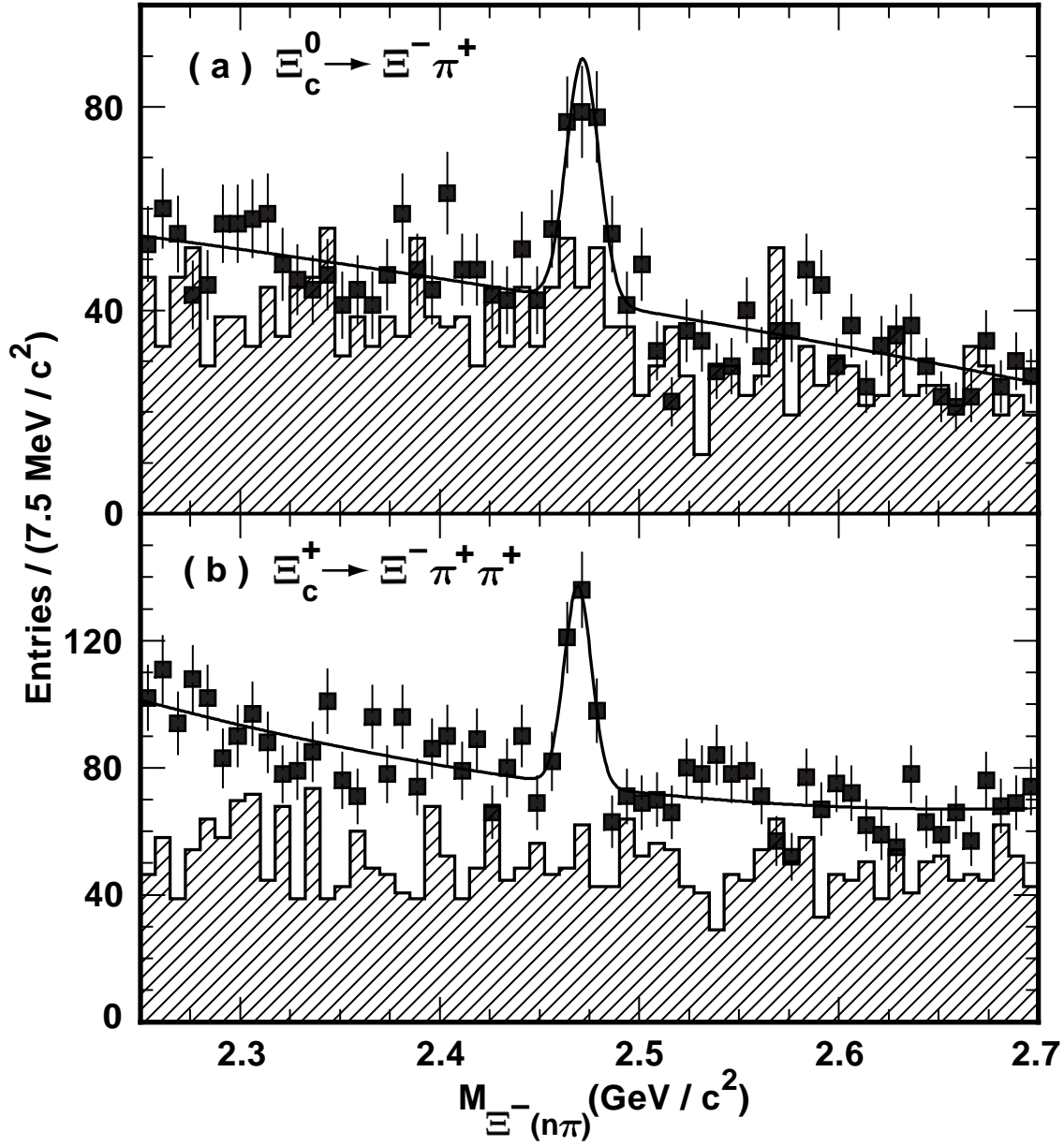}}
\vspace*{0.45in}
\caption{Invariant mass distributions of 
(a) \casm\pipl\ and (b) \casm\pipl\pipl\ 
from \ufours\ resonance (points) and scaled continuum (shaded histogram) data.} 
\label{fig:mass}
\end{figure}

To find the \casc\ signal yields, we fit each invariant mass distribution to the
sum of a Gaussian function of fixed width and a second order polynomial
background, both for the \ufours\ and the continuum data. 
The fixed widths for the two modes were determined  using a 
Monte Carlo simulation of the detector,
resulting in widths of 8.0 and 6.8 \mev\ 
for the \cascz\ and the \cascp, respectively.
We scale the continuum yields to account for the differences in 
luminosities and cross sections in the two data sets with the scale factor 
$({\cal L}_{\Upsilon(4S)}/{\cal L}_{\rm cont}) 
(E^2_{\rm cont}/E^2_{\Upsilon(4S)})$, where 
${\cal L}_{\Upsilon(4S)}$ and ${\cal L}_{\rm cont}$ are the luminosities, and
$ E_{\Upsilon(4S)}$ and $E_{\rm cont}$ are the beam energies
on the \ufours\ and on the continuum.
Figure~\ref{fig:mass} shows the invariant mass distributions 
of the \casm\pipl\ and \casm\pipl\pipl\ combinations 
from \ufours\ and scaled continuum data.
After subtracting the scaled continuum yield from the \ufours\ yield, 
we observe $79\pm 27$ \cascz\ candidates and 
$125\pm 28$ \cascp\ candidates from \B\ decays.
The errors are statistical only.
The fitted \casc\ masses are consistent with the current world averages.

To measure the product branching fractions
for the two decay modes, we divide both data and Monte Carlo
into $x_p$ intervals. The reconstruction
efficiency in each mode is found as a function of $x_p$ 
using Monte Carlo simulations.
Tables~\ref{table:cspi} and~\ref{table:cs2pip} show the continuum subtracted
raw yields $y_r(x_p)$ and efficiency-corrected yields $y_c(x_p)$. 
We also give the fractional decay rate in each $x_p$ interval, 
$(1/N_B)(dy_c/dx_p)$, where $N_B$ is $2 N_{B\overline{B}}$, 
for \cascz\ and \cascp\ production.
We find 
${\cal B}(\overline{B} \to \cascz\X){\cal B}(\cascz \to \casm\pipl) = 
(0.144\pm 0.048\pm 0.021)\times 10^{-3}$ and 
${\cal B}(\overline{B} \to \cascp\X){\cal B}(\cascp \to \casm\pipl\pipl) =
(0.453\pm 0.096\:^{+\:0.085}_{-\:0.065})\times 10^{-3}$, 
with the first error being statistical and the second being systematic.
The main sources of systematic error are due to uncertainties in the
reconstruction efficiencies for \lz\ (5\%) and \casm\ (7\%), 
variations in the selection criteria (8-9\%), uncertainties in 
particle identification (5\%), charged particle tracking (1\% per track), 
and the Monte Carlo predictions for the signal width (4\%). 
These result in a total systematic uncertainty of about 14\%.
In addition, we assign a $+12\%$ systematic uncertainty in the 
\casm\pipl\pipl\ case for the possible resonant substructure $\Xi^{*0}\pipl$, 
since this would decrease the \cascp\ reconstruction efficiency considerably.

We can convert these product branching fractions into absolute branching 
ratios using the following branching fractions of 
$\cascz \to \casm\pipl$ and $\cascp \to \casm\pipl\pipl$, derived by 
\cleo\cite{alexander}: 
${\cal B}(\cascz \to \casm\pipl)      = 
f_{SL} f_{\Xi_c} (0.52 \pm 0.16\:^{+\:0.15}_{-\:0.10})\%$ and
${\cal B}(\cascp \to \casm\pipl\pipl) = 
f_{SL} f_{\Xi_c} (2.5 \pm 0.6\:^{+\:0.8}_{-\:0.5})\%$,
where $f_{\Xi_c} \equiv {\cal B}(\casc\to\cas\ell^+\nul)/
{\cal B}(\casc\to\ell^+X) \leq 1$ (current predictions range from 
0.4 to 0.9\cite{kramer,bergfeld}), 
and $f_{SL} \equiv (\Gamma^{\Xi_c}_{SL} / \Gamma^{\Lambda_c}_{SL}) 
                   (\Gamma^{\Lambda_c}_{SL} / \Gamma^{D}_{SL})$, 
with $\Gamma_{SL}$ being the total semileptonic width.
These numbers are actually slightly different from the published values, since 
we are now using an updated value for 
$\Gamma^{D}_{SL} = (0.165 \pm 0.009)$~${\rm ps}^{-1}$\cite{pdg,dsem} 
(instead of the previous value of $(0.138 \pm 0.006)$~${\rm ps}^{-1}$). 
In addition, we have introduced the factor $f_{SL}$ to account for predictions 
of the semileptonic width of the \casc\ being quite different from that of 
the \lc\cite{voloshin} (2 to 3 times as large), which in turn should be 
different from that of the $D$\cite{bellini}, namely about 1.5 times as large. 
This leads to the following absolute branching ratios: 
${\cal B}(\overline{B} \to \cascz\X) = 
f_{SL}^{-1} f_{\Xi_c}^{-1}(2.8\pm 1.3\:^{+\:0.9}_{-\:0.7})\%$ and 
${\cal B}(\overline{B} \to \cascp\X) = 
f_{SL}^{-1} f_{\Xi_c}^{-1}(1.8\pm 0.6\:^{+\:0.7}_{-\:0.4})\%$.

\begin{figure}
\mbox{\psfig{figure=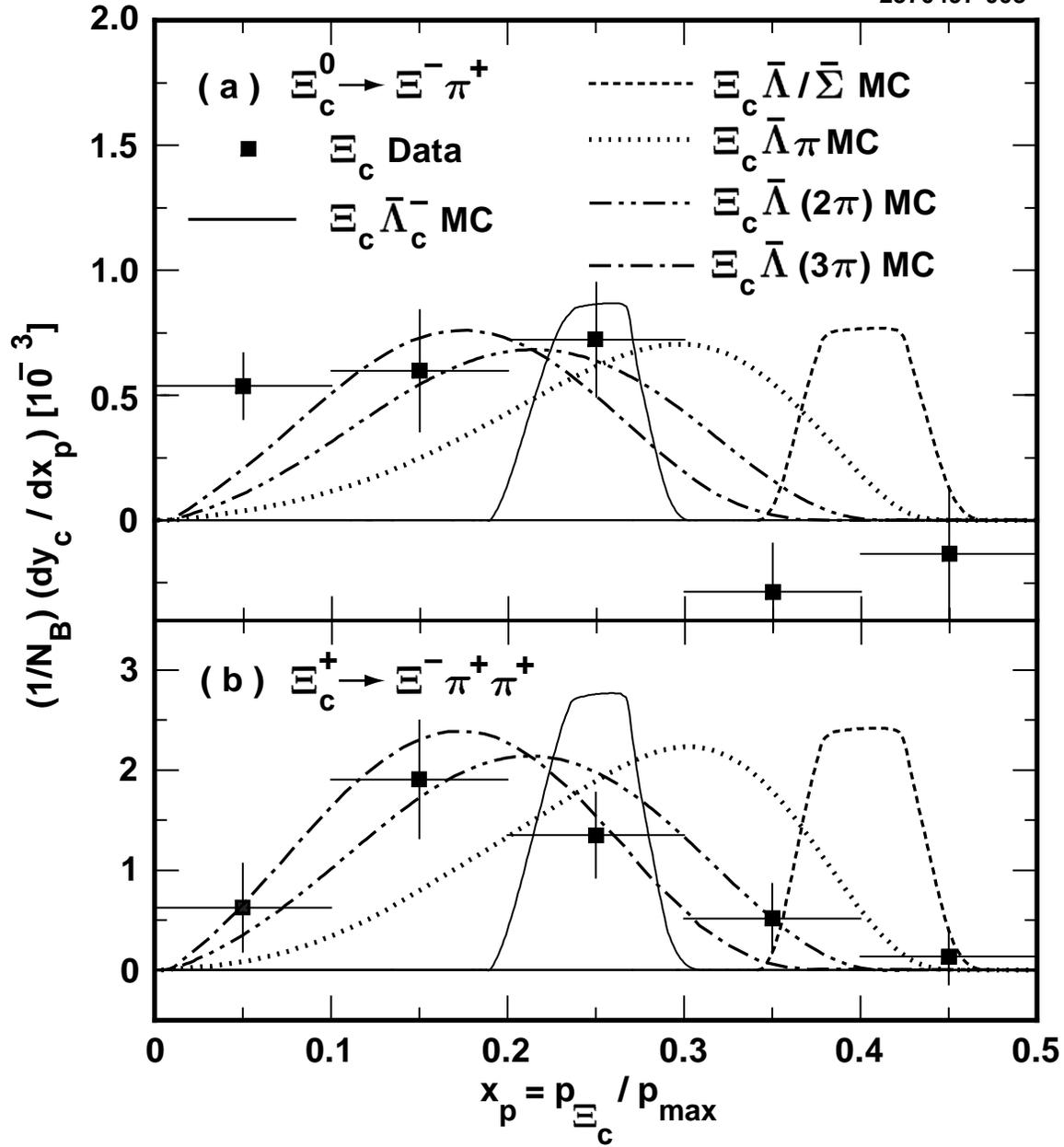}}
\vspace*{0.35in}
\caption{Efficiency-corrected momentum spectra for 
(a) \cascz\ and (b) \cascp\ from $B$ decays. 
The superimposed curves indicate the spectra derived from Monte Carlo
simulation of the decays 
$\overline{B} \to \casc\overline{\Lambda}_{(c)}(n\pi)$, $n=0,...,3$.
The Monte Carlo curves have been normalized to data, 
except for the two-body decays, where the normalization is arbitrary.}
\label{fig:xpover}
\end{figure}

In Figure~\ref{fig:xpover} we present 
the corresponding efficiency-corrected %and luminosity scaled
momentum spectra of \cascz\ and \cascp\ baryons in $B$ decays. 
Superimposed on the measured spectra are the results from 
Monte Carlo simulations of the decays 
$\overline{B} \to \casc\overline{\Lambda}_{(c)}(n\pi)$, $n=0,...,3$.
Comparing the measured spectra with Monte Carlo predictions indicates that 
two-body final states such as
$\casc\overline{\Lambda}$ and $\casc\overline{\Sigma}$
are suppressed while multi-body final states seem to be dominant. 
We are not yet sensitive to 
$b \to\ c\cbar s$ decays leading to final states of the form
$\casc\overline{\Lambda}_c$ or $\casc\overline{\Sigma}_c$, 
which are predicted by the authors of Refs.\cite{ball} and\cite{chernyak} 
to have branching fractions of only $(1.0-1.8)\times 10^{-3}$ for those decays. 
These branching fractions are at least an order of magnitude lower than 
the inclusive branching fractions for $\overline{B} \to \casc X$.

In summary, we have presented the first observation of \B\ mesons decaying into 
the charmed strange baryons \cascz\ and \cascp.
From an examination of the measured \cascz\ and \cascp\ momentum spectra, 
it is not clear which of the possible production mechanisms
$b \to c\ubar d$ or $b \to c\cbar s$ is preferred or dominant, 
since the observed momentum spectra are consistent with both mechanisms.
It seems, however, that decays involving a heavier anti-baryon or 
multi-body decays are favored.

We gratefully acknowledge the effort of the CESR staff in providing us with
excellent luminosity and running conditions.
This work was supported by 
the National Science Foundation,
the U.S. Department of Energy,
the Heisenberg Foundation,  
the Alexander von Humboldt Stiftung,
Research Corporation,
the Natural Sciences and Engineering Research Council of Canada,
and the A.P. Sloan Foundation.


 \begin{table}
 \begin{center}
 \caption{Inclusive \cascz\ production in B decays.}
 \label{table:cspi}
 \begin{tabular}{cr@{$\:\pm\:$}lr@{$\:\pm\:$}lr@{$\:\pm\:$}l}
% \hline \hline
 $\Delta \xp$ &\mcc{2}{Raw yield} 
 &\mcc{2}{Corr. yield} &\mcc{2}{$(1/N_B)(dy_c/d\xp)$} \\
 &\mcc{2}{$y_r(\xp)$} 
 &\mcc{2}{$y_c(\xp)$} &\mcc{2}{[$10^{-3}$]} \\ \hline %\hline
 $0.0-0.1$ &   27.0 &   6.5 &   358.8 &   88.1 &  0.54 & 0.13 \\
 $0.1-0.2$ &   33.4 &  13.5 &   399.5 &  162.3 &  0.60 & 0.24 \\
 $0.2-0.3$ &   43.5 &  13.6 &   482.8 &  152.5 &  0.72 & 0.23 \\
 $0.3-0.4$ &  -18.1 &  12.2 &  -191.5 &  129.5 & -0.29 & 0.19 \\
 $0.4-0.5$ &   -6.9 &  13.3 &   -89.7 &  174.1 & -0.13 & 0.26 \\
 \hline
 $0.0-0.5$ &   78.9 &  27.2 &   959.9 &  323.1 \\
% \hline \hline
 \end{tabular}
 \end{center}
 \end{table}

 \begin{table}
 \begin{center}
 \caption{Inclusive \cascp\ production in B decays.}
 \label{table:cs2pip}
 \begin{tabular}{cr@{$\:\pm\:$}lr@{$\:\pm\:$}lr@{$\:\pm\:$}l}
% \hline \hline
 $\Delta \xp$ &\mcc{2}{Raw yield} 
 &\mcc{2}{Corr. yield} &\mcc{2}{$(1/N_B)(dy_c/d\xp)$} \\
 &\mcc{2}{$y_r(\xp)$} 
 &\mcc{2}{$y_c(\xp)$} &\mcc{2}{[$10^{-3}$]} \\ \hline %\hline
 $0.0-0.1$ &   10.0 &   7.0 &   417.1 &  295.0 &  0.62 & 0.44 \\
 $0.1-0.2$ &   47.0 &  14.3 &  1273.5 &  392.6 &  1.91 & 0.59 \\
 $0.2-0.3$ &   41.8 &  13.0 &   901.4 &  285.5 &  1.35 & 0.43 \\
 $0.3-0.4$ &   20.2 &  13.6 &   344.2 &  232.8 &  0.52 & 0.35 \\
 $0.4-0.5$ &    6.0 &  12.4 &    89.6 &  186.0 &  0.13 & 0.28 \\
 \hline
 $0.0-0.5$ &  125.0 &  27.6 &  3025.8 &  641.5 \\
% \hline \hline
 \end{tabular}
 \end{center}
 \end{table}


\begin{references}
\bibitem{conj} Throughout this discussion, reference to a state also implies
reference to its charge conjugate state.
Also, it is assumed that both decays $\overline{B} \to \Xi_c X$ 
and $B \to \Xi_c X$ are possible.
\end{references}
\end{document}